\documentclass[aps,prl,twocolumn,superscriptaddress,nofootinbib]{revtex4-2}
\usepackage[T1]{fontenc}
\usepackage[utf8]{inputenc}
\usepackage{amsmath,amssymb,bm}
\usepackage{graphicx}
\usepackage[colorlinks=true,linkcolor=blue,citecolor=blue,urlcolor=blue]{hyperref}

\begin{document}

\title{Andreev Beta Conversion and Bulk Viscosity in Superfluid Neutron-Star Matter}

\author{Tingyu Zhang}
\affiliation{Department of Physics, State Key Laboratory of Optical Quantum Materials, and Hong Kong Institute of Quantum Science and Technology, University of Hong Kong, Hong Kong, China}

\author{Honoka Hoshino}
\affiliation{Department of Physics, Saga University, Saga 840-8502, Japan}

\author{Hiroyuki Tajima}
\affiliation{Department of Physics, Graduate School of Science, The University of Tokyo,
    Tokyo 113-0033, Japan}
\affiliation{RIKEN Nishina Center, Wako 351-0198, Japan}
\affiliation{Quark Nuclear Science Institute, The University of Tokyo, Tokyo 113-0033, Japan}

\author{Motoi Tachibana}
\affiliation{Department of Physics, Saga University, Saga 840-8502, Japan}
\affiliation{Center for Theoretical Physics, Khazar University, 41 Mehseti Street, Baku, AZ1096, Azerbaijan}

\author {Mariusz Sadzikowski}
\affiliation{Jagiellonian University, Institute of Theoretical Physics, \L{}ojasiewicza 11, 30-348 Kraków, Poland}

\date{\today}

\begin{abstract}
Bulk viscosity in neutron-star matter is attributed to the phase lag between a density oscillation and weak beta equilibration.  We point out a distinct coherent channel when the baryonic component that stores the condensate is superfluid: an effective beta-conversion event between a neutron condensate and normal protons can proceed as an Andreev process, converting a particle-like excitation into a hole-like excitation while adding or removing a Cooper pair from the condensate.  %
The resulting Andreev contribution to beta equilibration is fourth order in the effective conversion vertex, is enhanced by anomalous neutron Green's functions, and is not captured by standard Urca process.  We identify the density, temperature, and oscillation-frequency regimes where this channel can dominate viscous damping of neutron-star oscillations and mergers.
\end{abstract}

\maketitle

\textit{Introduction.---}
Bulk viscosity is one of the principal dissipative coefficients controlling the damping of neutron-star pulsations~\cite{PhysRevC.69.025802}, r modes~\cite{PhysRevLett.80.4843,PhysRevD.58.084020}, and post-merger density oscillations~\cite{PhysRevD.109.123022}. In nucleonic matter its standard origin is transparent: compression drives the system away from beta equilibrium, characterized by the chemical-potential imbalance $\delta\mu_\beta \equiv \mu_n-\mu_p-\mu_e$.
Weak reactions, such as $n\leftrightarrow p+e+\bar\nu_e$ restore equilibrium with a finite relaxation time, while the escaping neutrinos simultaneously provide an important cooling channel for compact stars~\cite{YAKOVLEV20011,yakovlev2004neutron}. The resulting pressure lag gives rise to a bulk viscosity $\zeta$ \cite{Sawyer1989,Haensel2000,Alford2018}.

The same density range is expected to contain paired baryons.  Neutrons can form a $^1S_0$ superfluid in the inner crust and outer core and a $^3P_2$ superfluid at higher density, while protons may be superconducting~\cite{RevModPhys.75.607,chamel2008physics,sedrakian2019superfluidity}.  Pairing suppresses ordinary quasiparticle Urca rates due to the pairing gap,
and standard calculations of superfluid bulk viscosity incorporate this suppression at the level of modified beta rates \cite{Haensel2000,Haensel2001}. 
On the other hand, the superfluidity opens an additional neutrino-emission channel through pair breaking and formation (PBF), 
which becomes important near the superfluid transition temperature but is suppressed at lower temperatures~\cite{1976ApJ,yakovlev1998neutrino}. Here we ask a different question: if normal-state protons coexist with superfluid neutrons, can beta conversion itself acquire an Andreev channel?

\begin{figure}[t]
    \centering
    \includegraphics[width=0.8\linewidth]{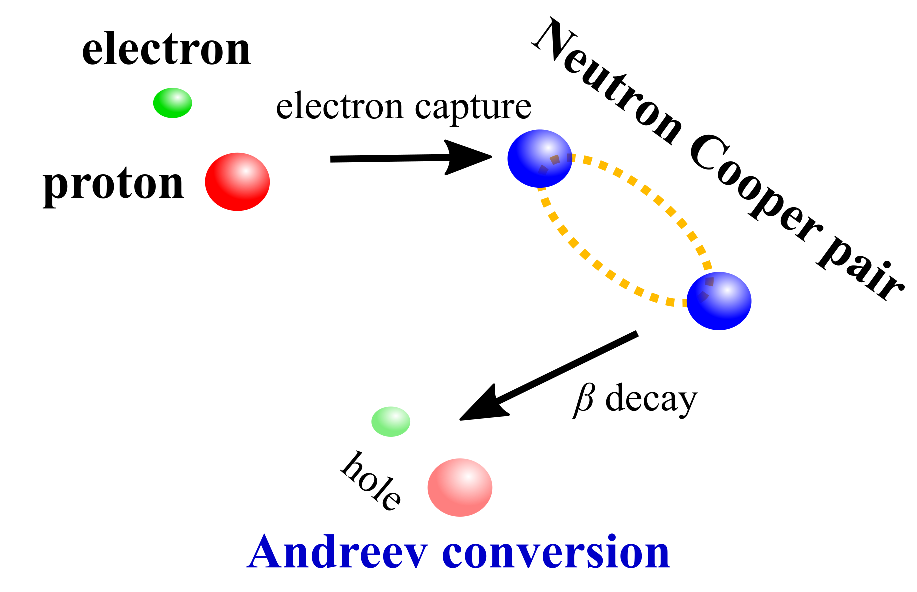}
    \caption{Schematics of the Andreev beta conversion between normal protons and neutron superfluidity. The formation of a neutron Cooper pair leads to the Andreev-type conversion of electron and proton holes.}
    \label{fig:1}
\end{figure}
Our motivation comes from a recent proposal and calculation of nonlinear radio-frequency transport in ultracold Fermi gases \cite{Zhang2023}.  There, a normal hyperfine state is coupled uniformly to a two-component Fermi superfluid.  Because the Rabi coupling conserves momentum, the fourth-order tunneling current contains a clean Andreev term governed by anomalous Green's functions.  At zero detuning the Andreev current can dominate over ordinary quasiparticle transport. On the other hand the Schwinger-Keldysh formalism based on non-equilibrium field theory provides a theoretical method to study the quantum transport in nuclear matter~\cite{7bj7-b15c,zhang2026nonequilibrium}. This allows us to study non-equilibrium astrophysical processes including cooling and density evolution in compact stars. Here we propose the neutron-star analogue shown schematically in Fig.~\ref{fig:1}:
the superfluid side is the neutron condensate, the normal component is the proton and electron fluids, and the Rabi coupling is replaced by an effective beta-conversion amplitude.  Neutrinos are not discarded physically; for the qualitative argument below they are integrated into the beta detuning and into a weak-conversion kernel. The question is then 
the dissipative relaxation of $\delta\mu_\beta$ in an oscillating star.

\textit{Effective model.---}
We separate the thermodynamic role of beta equilibrium from the coherent nucleonic channel.  
We consider the bulk Hamiltonian $H_0=H_n^{\rm SF}+H_p+H_e$,
where the neutrons are described in Nambu space by
\begin{equation}
  \mathcal{G}_n^{-1}(k)=
  \begin{pmatrix}
    i\omega_\ell-\xi_{n\bm{k}} & -\Delta_n(\bm{k})\\
    -\Delta_n^\ast(\bm{k}) & i\omega_\ell+\xi_{n\bm{k}}
  \end{pmatrix},
\end{equation}
where $\xi_{n,\bm{k}}=k^2/2m_n-\mu_n$, $\omega_\ell$ is the fermionic Matsubara frequency, and $\Delta_n(\bm{k})$ denotes the neutron pairing gap. The protons and electrons are treated as a normal Fermi liquid.
The beta tunnel Hamiltonian associated with the Fermi transition is written as
\begin{equation}
  H_T^{\rm eff}=
  \sum_{\bm{k}_p,\bm{k}_n,\bm{k}_e}
  \sum_{\sigma,\sigma_e}
  \left[
    g_F 
    \delta_{\bm{k}_n,\bm{k}_p+\bm{k}_e}
    p_{\bm{k}_p\sigma}^\dagger
    e_{\bm{k}_e\sigma_e}^\dagger
    n_{\bm{k}_n\sigma} 
    +{\rm h.c.}
  \right],
  \label{eq:HTeff}
\end{equation}
where $g_F$ is an effective beta-conversion vertex with dimension $[g_F]=E^{-1/2}$, where neutrinos are integrated out.

The net beta conversion rate is the tunneling current conjugate to $\delta\mu_\beta$, $\Gamma_{\beta}=-d\rho_n/dt$
with $\rho_n$ denoting the neutron number density.
Near the beta equilibrium ($\delta\mu_\beta\simeq 0$),
we obtain $\Gamma_{\beta}=\lambda\delta\mu_\beta+O((\delta\mu_\beta)^2)$, where $\lambda$ is the beta conversion conductance.
Expanding the Keldysh expectation value of $\Gamma_\beta$ in $H_T^{\rm eff}$, the second-order term gives the nucleon quasiparticle tunneling rate $\Gamma_{\rm qp}$, and its conductance $\lambda_{\rm qp}$, which is regarded as an analog of the quasiparticle conductance.
We obtain the second-order Keldysh expansion 
\begin{widetext}
\begin{align}
  \lambda_{\rm qp}
  =&
   |g_F|^2
  \sum_{\bm{k}_p,\bm{k}_n,\bm{k}_e}
  \sum_{\sigma_p,\sigma_e}
  \delta_{\bm{k}_n,\bm{k}_p+\bm{k}_e}
  \int
  \frac{d\omega_n}{2\pi}
  \frac{d\omega_p}{2\pi}
  \frac{d\omega_e}{2\pi}
  \,
  A_n(\bm{k}_n,\omega_n)
  A_p(\bm{k}_p,\omega_p)
  A_e(\bm{k}_e,\omega_e)
  \nonumber\\
  &\times
  2\pi\,
  \delta\!\left(
    \omega_n-\omega_p-\omega_e
  \right)
  \left[
    -\frac{\partial f(\omega_n)}{\partial \omega_n}
  \right]
  \bigl\{[1-f(\omega_p)]
  [1-f(\omega_e)]+f(\omega_p)f(\omega_e)\bigr\}.
  \label{eq:lambda_qp_df}
  \end{align}
\end{widetext}
where $A_i(\bm{k},\omega)=-2\,{\rm Im}\,G_i^R(\bm{k},\omega)$
is the spectral function of species $i=n,p,e$, and $f(\omega)$ is the Fermi distribution.  
The frequencies $\omega_{i=n,p,e}$ are measured
relative to the corresponding chemical potentials $\mu_{i}$.
In such a case, $\lambda_{\rm qp}$ is strongly suppressed due to the pairing gap in neutron superfluidity where the spectral function is gapped as
\begin{equation}
  A_n(\bm{k},\omega)
  =
  2\pi
  \left[
    u_{\bm{k}}^2 \delta(\omega-E_{\bm{k}})
    +
    v_{\bm{k}}^2 \delta(\omega+E_{\bm{k}})
  \right],
\end{equation}
with
  $E_{\bm{k}}
  =
  \sqrt{\xi_{n\bm{k}}^2+\Delta_n^2}$ and $u_{\bm{k}}^2=1-v_{\bm{k}}^2=(1+\xi_{n\bm{k}}/E_{\bm{k}})/2$.

The Andreev beta contribution appears at fourth order:
\begin{widetext}
\begin{align}
  \Gamma_A
  &=\frac{
  |g_F|^4}{2}
  \sum_{\bm{k}_p,\bm{k}_e}
  \sum_{\bm{k}'_p,\bm{k}'_e}
  \sum_{\sigma_p,\sigma_e,\sigma'_e}
  \delta_{\bm{k}_p+\bm{k}_e,\,-\bm{k}'_p-\bm{k}'_e}
  \int
  \frac{d\omega}{2\pi}
  \frac{d\omega_p}{2\pi}
  \frac{d\omega_e}{2\pi}
  \frac{d\omega'_p}{2\pi}
  \frac{d\omega'_e}{2\pi}
  \,
  \left|
    F_{n}(\bm{k}_p+\bm{k}_e,\omega)
  \right|^2
  \nonumber\\
  &\times
  A_p(\bm{k}_p,\omega_p)
  A_e(\bm{k}_e,\omega_e)
  A_p(-\bm{k}'_p,\omega'_p)
  A_e(-\bm{k}'_e,\omega'_e)
  (2\pi)^2\delta(
    \omega-\omega_p-\omega_e+\delta\mu_\beta
  )
  \delta(
    -\omega-\omega'_p-\omega'_e+\delta\mu_\beta
  )
  \nonumber\\
  &\times
  \Bigl[
    \{1-f(\omega_p)\}
    \{1-f(\omega_e)\}
    \{1-f(\omega'_p)\}
    \{1-f(\omega'_e)\}
    -
    f(\omega_p)
    f(\omega_e)
    f(\omega'_p)
    f(\omega'_e)
  \Bigr] .
  \label{eq:Gamma_A_beta_spectral}
\end{align}
\end{widetext}
where $F_n$ is the anomalous neutron Green's function, $A_p$ is the proton spectral function,
and
$A_e$ is the electron spectral function.
In contrast to $\Gamma_{\rm qp}$,
$\Gamma_A$ is not suppressed by the pairing gap as in the case of the Rabi-coupled Andreev reflection process~\cite{Zhang2023}.
The Andreev conductance is given by $\lambda_A=\big(\frac{\partial\Gamma_A}{\partial\delta\mu_\beta}\big)_{\delta\mu_\beta=0}$:
\begin{widetext}
\begin{align}
  \lambda_A
  &=\frac{
  |g_F|^4}{T}
  \sum_{\bm{k}_p,\bm{k}_e}
  \sum_{\bm{k}'_p,\bm{k}'_e}
  \sum_{\sigma_p,\sigma_e,\sigma'_e}
  \delta_{\bm{k}_p+\bm{k}_e,\,-\bm{k}'_p-\bm{k}'_e}
  \int
  \frac{d\omega}{2\pi}
  \frac{d\omega_p}{2\pi}
  \frac{d\omega_e}{2\pi}
  \frac{d\omega'_p}{2\pi}
  \frac{d\omega'_e}{2\pi}
  \,
  \left|
    F_{n}(\bm{k}_p+\bm{k}_e,\omega)
  \right|^2
  \nonumber\\
  &\times
  A_p(\bm{k}_p,\omega_p)
  A_e(\bm{k}_e,\omega_e)
  A_p(-\bm{k}'_p,\omega'_p)
  A_e(-\bm{k}'_e,\omega'_e)
  (2\pi)^2\delta(
    \omega-\omega_p-\omega_e
  )\delta(
    -\omega-\omega'_p-\omega'_e
  )
    f(\omega_p)
    f(\omega_e)
    f(\omega'_p)
    f(\omega'_e).
  \label{eq:Gamma_A_beta_spectral}
\end{align}
\end{widetext}
which dominates when the quasiparticle beta channel is suppressed by neutron pairing.  

\begin{table}[t]
    \caption{Proton fractions, Fermi momentum of neutrons, protons, and electrons, and the neutron gap energy for different baryon densities in a beta-equilibrium $npe$ matter.}
    \label{table1}
    \centering
    \begin{ruledtabular}
        \begin{tabular}{ccccc}
            $\rho_B/\rho_0$ & $x_p$ & $k_{Fn}$ [${\rm fm}^{-1}$] & $k_{Fp}=k_{Fe}$ [${\rm fm}^{-1}$] & $\Delta_n$ [MeV] \\
            0.1   & 0.0368 & 0.770  & 0.259 & 1.962 \\
            0.2  & 0.0285 & 0.973 & 0.300 & 1.434\\
            0.3 & 0.0268 & 1.114 & 0.337 & 0.776\\
            0.4 & 0.0268 & 1.226 & 0.370 & 0.285\\
        \end{tabular}
    \end{ruledtabular}
\end{table}
To study the Andreev and quasiparticle processes, we choose four different baryon densities for beta equilibrium $npe$ matters, and the corresponding proton fractions $x_p$ are given by Ref.~\cite{PhysRevLett.132.232701} at next-to-next-to-next-to-leading order (${\rm N}^3{\rm LO}$) level. For each density, the neutron gap energy $\Delta_n$ is obtained by a fitting formula~\cite{drissi2022many} to the numerical calculations using the chiral interaction at ${\rm N}^3{\rm LO}$~\cite{PhysRevC.68.041001} with a three-nucleon force at ${\rm N}^2{\rm LO}$. Table~\ref{table1} reports some physical quantities used for calculation obtained in different baryon densities, with $\rho_0\simeq 0.16\, {\rm fm}^{-3}$ denoting the nuclear saturation density.

According to results of numerical calculations~\cite{Supplement}, both the Andreev and quasiparticle contributions decrease as the temperature is lowered, but the latter exhibits a substantially stronger suppression. This qualitative difference can be understood from the nature of the two transport channels. The quasiparticle process requires thermally excited neutron Bogoliubov quasiparticles with excitation energy $E_{\bm{k}}=\sqrt{\xi^2_{n\bm{k}}+\Delta_n^2}\geq \Delta_n$. Therefore in low temperature regime $T\ll\Delta_n$, its phase space is exponentially suppressed: $\lambda_{\rm qp}\sim e^{-\Delta_n/T}$. In contrast, the Andreev process is mediated by the anomalous neutron propagator and represents a coherent pair-conversion process involving the condensate. It therefore does not require the thermal population of an on-shell neutron quasiparticle and is not subject to the same exponential suppression. Although $\lambda_A$ also decreases at low temperature because the available proton-electron phase space becomes progressively restricted, its decrease is much weaker than that of $\lambda_{\rm qp}$. These results suggest that the relative importance of the Andreev channel increases toward lower temperatures and that it may eventually dominate beta equilibration deep in the pairing regime.

\begin{figure}[t]
    \centering
    \includegraphics[width=1\linewidth]{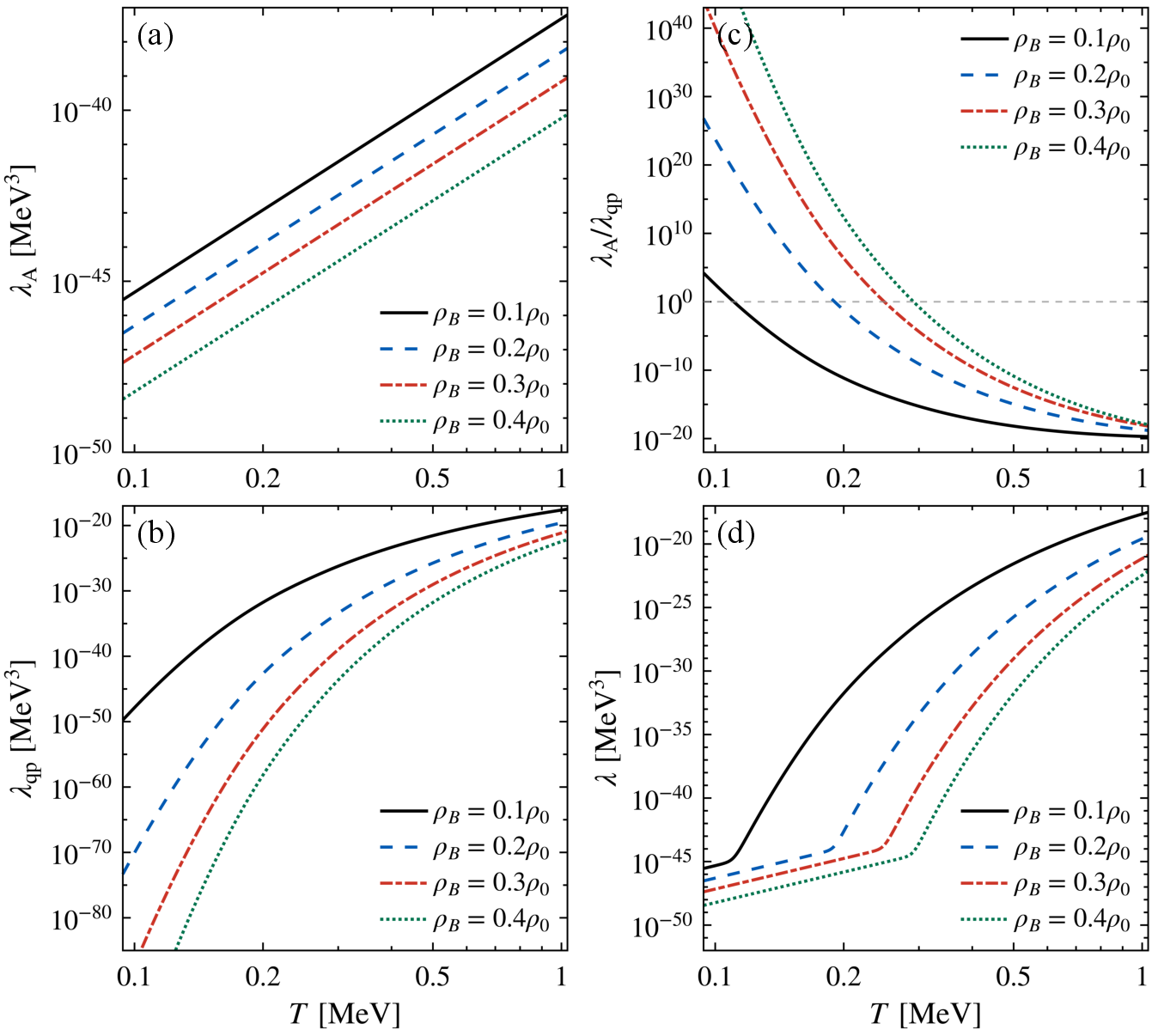}
    \caption{Asymptotic behavior of (a) Andreev conductance $\lambda_A$ and (b) quasiparticle conductance with varying temperatures $T$. (c) The ratio between $\lambda_A$ and $\lambda_{\rm qp}$. (d) The total conductance $\lambda=\lambda_A+\lambda_{\rm qp}$. We choose four different densities $\rho_B$ and proton fractions $x_p$ for beta equilibrium $npe$ matters, where $\rho_0\simeq 0.16\, {\rm fm}^{-3}$ is the saturation density.
    }
    \label{fig:lambda-T}
\end{figure}
To further understand the asymptotic behavior in extremely low temperatures, we use the approximation in BCS limit with $T\ll\Delta_n\ll \mu_n$, and the Fermi surface govern the total contribution of the $\beta$-conversion process~\cite{Supplement}. We found that in our cases where $k_{Fn}>2k_{Fp}$, the asymptotic expression of quasiparticle conductance can be given by
\begin{align}\label{asymp-qpconduct}
    \lambda_{\rm qp}\simeq\frac{2|g_F|^2k_{Fp}^3}{\pi^3 v_{Fp}}\frac{E_*^2}{|\xi_*|v_*}e^{-E_*/T}.
\end{align}
Here we define $E_*=E_{2k_{Fp}}$ and $\xi_*=\xi_{n}(2k_{Fp})$. $v_{Fp}=k_{Fp}/\sqrt{k_{Fp}^2+m^2}$ is the Fermi velocity of protons, and $v_*=2k_{Fp}/\sqrt{(2k_{Fp})^2+m^2}$. By comparing our asymptotic expression in $\Delta_n=0$ case with a typical result of direct Urca conductance~\cite{9nck-dm66}, the expression of the effective coupling vertex is given by
\begin{align}
    |g_F|^2\simeq \frac{17}{60}(1+3g_A^2)\cos^2{\theta_c}G_F^2T^3,
\end{align}
where $\theta_c$ is the Cabibbo angle and $g_A=1.26$ is the axial vector coupling constant.
The asymptotic behavior of Andreev conductance can be obtained as 
\begin{align}\label{asymp-Andreevconduct}
    \lambda_{A}\simeq\frac{|g_F|^4k^4_{Fp}}{9\pi^3v_{Fp}^2v_*}\frac{\Delta_n^2T^2}{|\xi_*|^3}.
\end{align}
According to Eqs~(\ref{asymp-qpconduct}) and (\ref{asymp-Andreevconduct}), as well as the $T$-dependence of $g_F$, at low temperatures $T\ll \Delta_n$, the quasiparticle conductance and Andreev conductance respectively behave as $\lambda_{\rm qp}\propto T^3e^{-E_*/T}$ and $\lambda_A\propto {\Delta_n^2T^8}/{|\xi_*|^3}$. Therefore, when the temperature becomes sufficiently lower than the gap energy, $\lambda_{\rm qp}$ decreases much more rapidly than $\lambda_{A}$, as shown in Fig.~\ref{fig:lambda-T}. This is consistent with our numerical results shown in the supplementary material~\cite{Supplement}. To compare the importance of two contributions at different temperatures we plot the asymptotic results of the ratio $\lambda_{A}/\lambda_{\rm qp}$ and total conductance $\lambda=\lambda_{\rm qp}+\lambda_{A}$ in Fig.~\ref{fig:lambda-T} (c) and (d). 
Although $\lambda_A$ and $\lambda_{\rm qp}$ differ by a factor $|g_F|^2$, at sufficiently low temperature, e.g., $T\sim 0.1\,{\rm MeV}$, the Andreev contribution becomes more important than the quasiparticle one and dominate the whole conversion process.
In the plot of total conductance, the flat part of the curve represents an Andreev-dominated conductance, while the non-linear part represents a quasiparticle-dominated one.

We note that the low-temperature behavior discussed above depends on the structure of neutron pairing gap.
Although the present calculation assumes an isotropic $S$-wave gap, 
an anisotropic $P$-wave pairing
may occur at higher densities~\cite{PhysRevLett.24.775,tamagaki1970superfluid} or under a strong magnetic field~\cite{PhysRevC.108.L052802}. 
In particular, nodal pairing states contain gapless directions on the neutron Fermi surface, which qualitatively modifies the low-temperature asymptotics.
More specifically, for $k_{Fn}<2k_{Fp}$, the gapless $P$-wave pairing gives $\lambda_{\rm qp}\propto T^6$~\cite{Haensel2000,1994AstL}.
Thus, the relative importance of the Andreev and quasiparticle channels can depend on the pairing symmetry.

\textit{Bulk viscosity.---}
For a baryon density perturbation $\rho_B(t)=\bar \rho_B+\delta \rho_B e^{i\omega t}$ (where $\rho_B=\rho_n+\rho_p$ and $\bar{\rho}_B$ is its equilibrium value) with a frequency $\omega$, the beta imbalance $\delta\mu_\beta(t)$ obeys the linearized equation
\begin{equation}
\label{eq:14}
  \frac{d}{dt}\delta{\mu}_\beta(t)
  = \,\frac{C}{\bar\rho_B}\frac{d\rho_B}{dt}
  - B\,\Gamma_\beta,
\end{equation}
where $B=-\frac{1}{\rho_B}\left(\frac{\partial\delta\mu_\beta}{\partial x}\right)_{\rho_B}$ and $C=\rho_B\left(\frac{\partial\delta\mu_\beta}{\partial\rho_B}\right)_x$ are susceptibilities obtained from the EoS. Using an appropriate symmetry energy $S(\rho_B)$ incorporating baryons interactions~\cite{PhysRevC.104.055805,PhysRevD.105.043015,sty1706,PhysRevC.58.1804}, we have $\mu_n-\mu_p=4S(\rho_B)(1-2x_p)$. Then using $\mu_e=(3\pi^2\rho_B x_p)^{1/3}$, we obtain
\begin{align}
    &B=\frac{\mu_e}{\rho_B}\left(\frac{2}{1-2x_p}+\frac{1}{3x_p}\right)\ ,\nonumber\\
    &C=4\rho_B(1-2x_p)\frac{dS}{d\rho_B}-\frac{\mu_e}{3}=\rho_B^2B\frac{dx_p}{d\rho_B}.
\end{align}

Using $\delta\mu_{\beta}(t)=\delta\tilde{\mu}_\beta(\omega)e^{i\omega t}$,
we obtain
\begin{align}
\label{eq:15}
    \delta\tilde{\mu}_\beta(\omega)=\frac{i\omega C}{i\omega+B\lambda}\frac{\delta\rho_B}{\bar\rho_B},
\end{align}
where $\lambda=\lambda_{\rm qp}+\lambda_A$.


The dynamical bulk viscosity $\zeta(\omega)$ is introduced as
\begin{align}
\tilde{\Pi}(\omega)=-\zeta(\omega)\nabla\cdot\tilde{\bm{v}}(\omega), 
\end{align}
where $\tilde{\Pi}$ is the Fourier component of the instantaneous nonequilibrium pressure $\Pi=P-\bar{P}$ measured from the equilibrium one $\bar{P}$ with fixed ${\rho}_B$, and $\nabla\cdot\tilde{\bm{v}}(\omega)=-\frac{i\omega}{\bar{\rho}_B}\delta\rho_B.$ is the Fourier component
of $\nabla\cdot\tilde{\bm{v}}$. 
The analytical formula of ${\rm Re}\,\zeta(\omega)$ is given by
\begin{equation}
  {\rm Re}\,\zeta(\omega) =
  \frac{C^2\lambda}
       {\omega^2+(B\lambda)^2},
  \label{eq:zeta}
\end{equation}

\begin{figure}[t]
    \centering
    \includegraphics[width=1\linewidth]{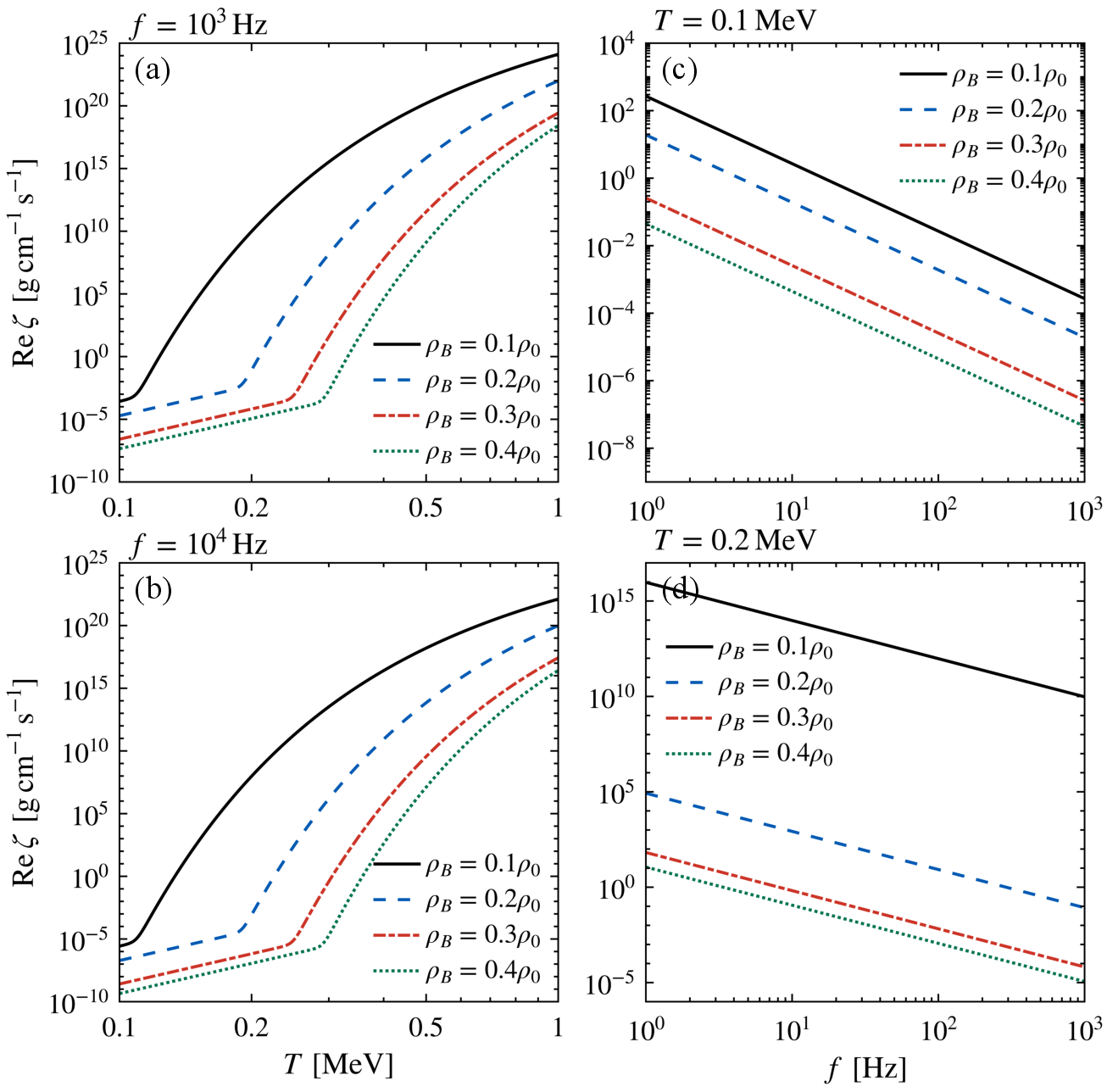}
    \caption{The Real part of dynamical bulk viscosity $\operatorname{Re}\zeta(\omega)$ arising from beta conversions, as a function of (a)-(b) temperature $T$, and (c)-(d) frequency $f=\omega/2\pi$. The results are obtained based on the asymptotic expressions of $\lambda$ at low temperatures, and are evaluated within three different oscillation frequencies $\omega$, and four different baryon densities. Notice that in panel (d) the black curve is dominated by the quasiparticle contribution, while all the others are genuine Andreev effect.
    }
    \label{fig:zeta}
\end{figure}
Fig.~\ref{fig:zeta} shows $\operatorname{Re}\zeta(\omega)$ as functions of temperature $T$ and frequency $f=\omega/2\pi$, at different baryon densities. The low-temperature response is governed by condensate-mediated Andreev conversion and exhibits a dependence as $\operatorname{Re}\zeta\propto T^8$, rather than the exponential suppression characteristic of the quasiparticle channel. For the adopted microscopic and thermodynamic inputs, increasing the baryon density reduces the viscosity and shifts this crossover toward higher temperatures. Importantly, although Andreev conversion dominates the low-temperature beta response, the absolute viscosity continues to decrease upon cooling.

\textit{Discussion.---}
Unlike superfluid suppression factors that modify ordinary quasiparticle rates, Andreev beta conversion involves coherent anomalous propagation through the neutron condensate. It contributes to composition relaxation in multifluid hydrodynamics, where entrainment and relative fluid motion may affect its role.
This mechanism also differs from phonon collisions, which relax the collective sector through strong interactions and can dominate bulk viscosity near $T\sim 10^9\ {\rm K}$ in some models studied~\cite{Cristina_Manuel_2013,universe7030059}. Andreev conversion instead requires coexistence of superfluid neutrons and normal protons, and relaxes the composition imbalance $\delta\mu_{\beta}$ through weak interactions.

Goldstone modes also contribute to cooling, with estimated emissivity scalings $\mathcal{E}_{\rm ph} \propto G_F^2 T^{17}$ for phonons and $\mathcal{E}_{\rm ang} \propto G_F^2T^9$ for angulons in the $^3P_2$ phase~\cite{universe7030059}. In contrast, the Andreev conductance $\lambda_A\propto G_F^4 T^8$ suggests an emissivity imbalance response $\partial\delta\mathcal{E}_A/\partial\delta\mu\sim T\lambda_A\propto G_F^4T^9$, where $\delta\mathcal{E}_A$ is the difference between antineutrino and neutrino energy emissivities. Assuming characteristic neutrino energies of order $T$, phase-space counting further suggests a total estimated equilibrium emissivity $\mathcal{E}_{A,0}\sim T^2\lambda_A\propto G_F^4T^{10}$, built on the calculated conductance rather than an explicit emissivity calculation. Therefore, despite its higher weak-coupling order, Andreev emission may provide an important low-temperature cooling channel in fully gapped phases, potentially exceeding the phonon emission. While precision laboratory experiments probe electroweak interactions with high accuracy~\cite{2006257,PhysRevLett.106.041803,jefferson2018precision}, here our results suggest an unexpected potential astrophysical probe of higher-order weak processes through condensate-mediated transport and cooling of the neutron stars. 

\textit{Conclusion.---}
We have discussed an Andreev beta conversion and its impact on the bulk viscosity in superfluid neutron-star matter.  By
using the Schwinger--Keldysh approach for calculating the higher-order tunneling currents~\cite{Zhang2023} with a reduced nucleonic beta-conversion Hamiltonian, we find a fourth-order anomalous Andreev beta relaxation channel governed by neutron condensate Green's functions.
At low temperature, the Andreev process becomes dominant even compared to the second-order quasiparticle-conversion process.
Our results indicate that the Andreev process plays a crucial role in the neutron-star dynamics.

For future perspectives, 
it is worthwhile to estimate
the effective amplitude $g_F$ by including lepton phase space and Fermi-liquid renormalizations. 
While we consider the momentum conservation of the effective vertex,
it would require more careful consideration in the inhomogeneous inner crust or vortex-lattice environment. 

\begin{acknowledgments}
H.~T. thanks H. Liang for useful discussions.
T.~Z. acknowledges support from HK GRF Grants No.~17306024, CRF Grants No.~C7015-24GF and C4050-23GF, CRS\_HKU701/24, and a RGC Fellowship Award No. HKU RFS2223-7S03.
\end{acknowledgments}

\bibliography{ref.bib}

\clearpage

\begin{widetext}

\begin{center}
{\Large \textbf{Supplemental Material for \\ ``Andreev Beta Conversion and Bulk Viscosity in Superfluid Neutron-Star Matter"}}
\end{center}

\section{Asymptotic behavior of conductance at low temperatures}

Here we derive the asymptotic expression for conductances of the quasiparticle and the Andreev beta conversion at extremely low temperatures where $T\ll\Delta_n\ll\mu_n$. 

\subsection{Quasiparticle Conductance}
According to Eq.~(9) in the main text, after the frequency integral, $\lambda_{\rm qp}$ has the form:
\begin{align}\label{eq:qpconduct}
    \lambda_{\rm qp}=8\pi|g_{F}|^2\int\frac{d^3\bm{q}}{(2\pi)^3}\int\frac{d^3 \bm{k}}{(2\pi)^3}\bigg\{&u^2_{\bm{q}}\,\delta\big[E_{\bm{q}}-\xi_p(\bm{k})-\xi_e(\bm{q}-\bm{k})\big]\frac{e^{E_{\bm{q}}/T}}{T(e^{E_{\bm{q}}/T}+1)^2}S_+\nonumber\\
    +&v_{\bm{q}}^2\delta\big[-E_{\bm{q}}-\xi_p(\bm{k})-\xi_e(\bm{q}-\bm{k})\big]\frac{e^{E_{\bm{q}}/T}}{T(e^{E_{\bm{q}}/T}+1)^2}S_-\bigg\},
\end{align}
with $S_+=[1-f(\xi_p(\bm{k}))][1-f(E_{\bm{q}}-\xi_p(\bm{k}))]+f(\xi_p(\bm{k}))f(E_{\bm{q}}-\xi_p(\bm{k}))$ and $S_-=[1-f(\xi_p(\bm{k}))][1-f(-E_{\bm{q}}-\xi_p(\bm{k}))]+f(\xi_p(\bm{k}))f(-E_{\bm{q}}-\xi_p(\bm{k}))$.
In the BCS limit, We use the DOS approximation where the Fermi surface gives all contribution to the conversion. Due to the momentum conservation $\bm{q}=\bm{k}_p+\bm{k}_e$, we have
\begin{align}
    \int\frac{d^3\bm{k}}{(2\pi)^3}\rightarrow\int\frac{d^3\bm{k}}{(2\pi)^3}\delta\big[\xi_p(\bm{k})\big]\delta\big[\xi_e(\bm{q}-\bm{k})\big]\int_{-\infty}^\infty d\xi_p \int_{-\infty}^\infty d\xi_e.
\end{align}
For isotropic dispersion, we define
\begin{align}\label{eq:Iq}
    \mathcal{I}(q)= &\int\frac{d^3\bm{k}}{(2\pi)^3}\delta\big[\xi_p(k)\big]\delta\big[\xi_e(|\bm{q}-\bm{k}|)\big]\nonumber\\
    =&\frac{k_{Fp}k_{Fe}}{4\pi^2qv_{Fp}v_{Fe}}\Theta(q-q_-)\Theta(q_+-q),
\end{align}
where $q_-=|k_{Fp}-k_{Fe}|$, $q_+=k_{Fp}+k_{Fe}$, $v_{Fp}=k_{Fp}/\sqrt{k_{Fp}^2+m^2}$, and $v_{Fe}\simeq 1$. Additionally, to maintain charge neutrality, we always have $k_{Fp}=k_{Fe}$. Therefore Eq.~(\ref{eq:Iq}) can be rewritten as 
\begin{align}
    \mathcal{I}(q)=\frac{k^2_{Fp}}{4\pi^2qv_{Fp}}\Theta(2k_{Fp}-q).
\end{align}
Using the equality
\begin{align}
    \int_{-\infty}^{\infty}d\xi \big[(1-f(\xi))(1-f(\Omega-\xi))+f(\xi)f(\Omega-\xi)\big]=\Omega\coth{\frac{\Omega}{2T}},
\end{align}
we can then rewrite Eq.~(\ref{eq:qpconduct}) as
\begin{align}\label{eq:qpconduct-BCS}
    \lambda_{\rm qp}=&8\pi|g_{F}|^2\int\frac{d^3\bm{q}}{(2\pi)^3}\mathcal{I}(q)\frac{e^{E_{\bm{q}}/T}}{T(e^{E_{\bm{q}}/T}+1)^2}E_{\bm{q}}\coth{\frac{E_{\bm{q}}}{2T}}\nonumber\\
    =&4\pi|g_{F}|^2\int\frac{d^3\bm{q}}{(2\pi)^3}\mathcal{I}(q)\frac{E_{\bm{q}}}{T\sinh(E_{\bm{q}}/T)}.
\end{align}

At low temperature limit $T\ll \Delta_n$, Eq.~(\ref{eq:qpconduct-BCS}) becomes
\begin{align}\label{eq:qpconduct-lowT}
    \lambda_{\rm qp}=8\pi|g_{F}|^2\int\frac{d^3\bm{q}}{(2\pi)^3}\mathcal{I}(q)\frac{E_{\bm{q}}}{T}e^{-E_{\bm{q}}/T}.
\end{align}
Next, we will discuss the magnitude of the current by considering two different cases. First, if $k_{Fn}\leq 2k_{Fp}$, the most contribution lies on the Fermi surface of neutrons: $q=k_{Fn}$. We can thus assume that
\begin{align}
    \lambda_{\rm qp}\simeq8\pi|g_{F}|^2N_n(0)\mathcal{I}(k_{Fn})\int_{-\infty}^{\infty} d\xi_n\frac{E}{T}e^{-E/T},
\end{align}
where $N_n(0)=k_{Fn}^2/(2\pi^2v_{Fn})$ is the density of
states at the neutron Fermi surface. Let $\xi_n=\Delta_n\sinh{u}$, $E=\Delta_n\cosh{u}$, $d\xi_n=\Delta_n\cosh{u}\,du$. Then
\begin{align}
    \int_{-\infty}^{\infty} d\xi_n\frac{E}{T}e^{-E/T}=\frac{\Delta_n^2}{T}\int_{-\infty}^{\infty} du\cosh^2{u}e^{-\Delta_n\cosh{u}/T}.
\end{align}
Since $\Delta_n/T\gg 1$, the interval where $u\simeq 0$ dominates the above integral, and $\cosh{u}\simeq 1+u^2/2$. Therefore Eq.~(\ref{eq:qpconduct-lowT}) can be rewritten as
\begin{align}
    \lambda_{\rm qp}\simeq&8\pi\sqrt{2\pi}|g_F|^2N_n(0)\mathcal{I}(k_{Fn})\Delta_n\sqrt{\frac{\Delta_n}{T}}e^{-\Delta_n/T}\nonumber\\
    =&\frac{\sqrt{2\pi}|g_F|^2k_{Fn}k^2_{Fp}}{\pi^3v_{Fn}v_{Fp}}\Delta_n\sqrt{\frac{\Delta_n}{T}}e^{-\Delta_n/T},
\end{align}
which is proportional to $e^{-\Delta_n/T}$.

If $k_{Fn}>2k_{Fp}$, which happens in neutron rich beta equilibrium $npe$ matters with baryon density below $\rho_0$, the Fermi surface of neutron does not contribute to the beta conversion process due to the momentum conservation. The quasiparticle can then be written as 
\begin{align}\label{eq:qpconduct-case2}
    \lambda_{\rm qp}\simeq\frac{|g_F|^2k^2_{Fp}}{\pi^3v_{Fp}}\int_0^{2k_{Fp}}dq\frac{qE_{\bm{q}}}{T}e^{-E_{\bm{q}}/T}.
\end{align}
Since in the interval $0<q<2k_{Fp}$, $E_{\bm{q}}$ decreases as $q$ increases, at low temperature the integral above is dominated by the endpoint $q\simeq 2k_{Fp}$. Defining $E_*=\sqrt{\xi_n(2k_{Fp})^2+\Delta_n^2}$, near the endpoint, $E_{\bm{q}}\simeq E_*+|E'_*|\delta q$. Therefore Eq.~(\ref{eq:qpconduct-case2}) can be rewritten as 
\begin{align}
    \lambda_{\rm qp}\simeq&\frac{|g_F|^2k^2_{Fp}}{\pi^3v_{Fp}}\frac{2k_{Fp}E_*}{|E_*'|}e^{-E_*/T}\nonumber\\
    =&\frac{2|g_F|^2k_{Fp}^3}{\pi^3 v_{Fp}}\frac{E_*^2}{|\xi_*|v_*}e^{-E_*/T},
\end{align}
which is our case. Here $\xi_*=\xi_{n}(2k_{Fp})$, and $v_*=2k_{Fp}/\sqrt{(2k_{Fp})^2+m^2}$.

\subsection{Temperature Dependence of $g_F$}

According to Eq.~\ref{eq:qpconduct-BCS}, for normal (unpairing) neutron, $\Delta_n=0$, the conductance of quasiparticle $\beta$ conversion becomes
\begin{align}
    \lambda_{\rm qp}=4\pi|g_F|^2N_n(0)I(k_{Fn})\int d\xi\frac{|\xi|}{T\sinh(|\xi|/T)}.
\end{align}
Since 
\begin{align}
\int_{-\infty}^{\infty}
d\xi\,\frac{|\xi|}{T\sinh\left(|\xi|/T\right)}=\frac{\pi^2}{2}T,
\end{align}
we have
\begin{align}
    \lambda_{\rm qp}=\frac{|g_F|^2}{4\pi}\frac{k_{Fn}k_{Fp}k_{Fe}}{v_{Fn}v_{Fp}v_{Fe}}T.
\end{align}
Using $k_{Fi}/v_{Fi}=E_{Fi}$, it becomes
\begin{align}\label{normal-conductance}
    \lambda_{\rm qp}=&\frac{|g_F|^2}{4\pi}E_{Fn}E_{Fp}E_{Fe}T\nonumber\\
    \simeq&\frac{|g_F|^2}{4\pi}E_{Fn}E_{Fp}k_{Fe}T,
\end{align}
where $E_{Fi}=\sqrt{k_{Fi}^2+m^2}$ is the relativistic Fermi energy of baryons. 

On the other hand, a typical result of the linear direct-Urca rate obtained by relativistic Fermi-surface calculation~\cite{9nck-dm66} is given by
\begin{align}
    \lambda_{\mathrm{DU}}=\frac{17}{240\pi}G_F^2\cos^2\theta_c\left(1+3g_A^2\right)E_{Fn}E_{Fp}k_{Fe}T^4,
\end{align}
where $\theta_c$ is the Cabibbo angle and $g_A=1.26$ is the axial vector coupling constant. Comparing this expression with Eq.~(\ref{normal-conductance}), we can see that
\begin{align}
    |g_F|^2\simeq \frac{17}{60}(1+3g_A^2)\cos^2{\theta_c}G_F^2T^3.
\end{align}
Therefore actually $g_F$ is rather dependent on temperature:
\begin{align}
    g_F\propto G_F T^{3/2}. 
\end{align}

\subsection{Andreev Conductance}

Next we derive the asymptotic formula for the conductance of Andreev beta conversion, $\lambda_A$, using the same approximation. We define 
\begin{align}
    \mathcal{P}(\bm{q},\omega)=&\int\frac{d^3\bm{k}}{(2\pi)^3}\delta[\omega-\xi_p(\bm{k})-\xi_e(\bm{q}-\bm{k})]f[\xi_p(\bm{k})]f[\xi_e(\bm{q}-\bm{k})]\nonumber\\
    \simeq&\mathcal{I}(q)\int_{-\infty}^{\infty}d\xi f(\xi)f(\omega-\xi)\nonumber\\
    =&\mathcal{I}(q)\frac{\omega}{e^{\omega/T}-1}.
\end{align}
After frequency integration over delta functions, the Andreev conductance can be rewritten as
\begin{align}\label{eq:Andreevconduct-BCS}
    \lambda_A=&\frac{16\pi|g_F|^4}{T}\int\frac{d^3\bm{q}}{(2\pi)^3}\int_{-\infty}^{\infty} d\omega\big|F_n(\bm{q},\omega)\big|^2\mathcal{P}(\bm{q},\omega)\mathcal{P}(\bm{q},-\omega)\nonumber\\
    =&\frac{4\pi|g_F|^4}{T}\int\frac{d^3\bm{q}}{(2\pi)^3}\int_{-\infty}^{\infty} d\omega\big|F_n(\bm{q},\omega)\big|^2\mathcal{I}^2(q)\frac{\omega^2}{\sinh^2{(\omega/2T)}}.
\end{align}
In low temperature limit, the $\frac{\omega^2}{\sinh^2{(\omega/2T)}}$ in the integral restricts the frequency in the interval where $|\omega|\sim T$. Moreover, since $T\ll \Delta_n\leq E_{\bm{q}}$, the anomalous Green's function of neutrons can be approximated as $F_n\simeq-\Delta_n/E^2_{\bm{q}}$. Using
\begin{align}
    \int_{-\infty}^{\infty}d\omega \frac{\omega^2}{\sinh^2{(\omega/2T)}}=\frac{8\pi^2}{3}T^3,
\end{align}
Eq.~(\ref{eq:Andreevconduct-BCS}) can be rewritten as 
\begin{align}
    \lambda_A\simeq\frac{32\pi^3|g_F|^4T^2}{3}\int\frac{d^3\bm{q}}{(2\pi)^3}\mathcal{I}^2(q)\frac{\Delta_n^2}{E_{\bm{q}}^4}.
\end{align}
If $k_{Fn}<2k_{Fp}$, like the quasiparticle case, the momentum integration can be approximated as integration over DoS at the neutron Fermi surface:
\begin{align}
    \lambda_A\simeq&\frac{32\pi^3|g_F|^4}{3}T^2N_n(0)\mathcal{I}^2(k_{Fn})\int d\xi_n\frac{\Delta_n^2}{(\xi_n^2+\Delta_n^2)^2}\nonumber\\
    =&\frac{16\pi^3|g_F|^4}{3}N_n(0)\mathcal{I}^2(k_{Fn})\frac{T^2}{\Delta_n},
\end{align}
which is proportional to $T^2/\Delta_n$.

\begin{figure}[t]
    \centering
    \includegraphics[width=0.95\linewidth]{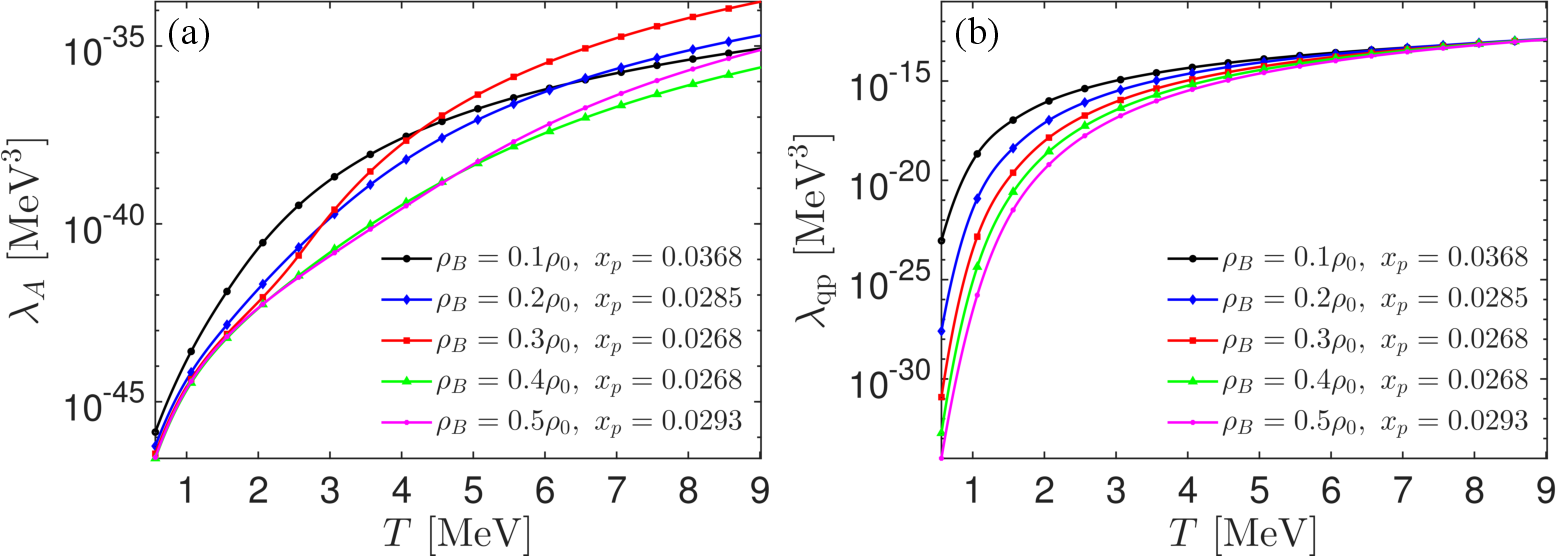}
    \caption{Numerical results of (a) Andreev conductance $\lambda_A$ and (b) quasiparticle conductance as functions of temperature $T$. We choose five different densities $\rho_B$ and proton fractions $x_p$ for beta equilibrium $npe$ matters, where $\rho_0\simeq 0.16\, {\rm fm}^{-3}$ is the saturation density.
    }\label{fig:num-lambda-T}
\end{figure}
On the other hand, if $k_{Fn}>2k_{Fp}$,
\begin{align}\label{eq:Andreevconduct-case2}
    \lambda_A\simeq\frac{|g_F|^4k_{Fp}^4}{3\pi^3v^2_{Fp}}T^2\int_0^{2k_{Fp}}dq\frac{\Delta_n^2}{[\xi_n(q)^2+\Delta_n^2]^2}.
\end{align}
Near the endpoint of the integral, $\xi_n(q)\simeq\xi_*+v_*\delta q$, with $\xi_*<0$. Therefore
\begin{align}
    \int_0^{2k_{Fp}}dq\frac{\Delta_n^2}{[\xi_n(q)^2+\Delta_n^2]^2}\simeq&\int_0^\infty dy\frac{\Delta_n^2}{[(|\xi_*|+v_*y)^2+\Delta_n^2]}\nonumber\\
    =&\frac{1}{v_*}\int_{\xi_*}^\infty du\frac{\Delta_n^2}{(u^2+\Delta_n^2)^2}\nonumber\\
    =&\frac{1}{v_*}\bigg[\frac{\pi}{4\Delta_n}-\frac{|\xi_*|}{2(|\xi_*|^2+\Delta_n^2)}-\frac{1}{2\Delta_n}\arctan{\frac{|\xi_*|}{\Delta_n}}\bigg]\nonumber\\
    =&\frac{1}{2v_*\Delta_n}\arctan{\frac{\Delta_n}{|\xi_*|}}-\frac{|\xi_*|}{2v_*(|\xi_*|^2+\Delta_n^2)},
\end{align}
where we let $u=|\xi_*|+v_*y$, $dy=du/v_*$. Then Eq.~(\ref{eq:Andreevconduct-case2}) can be rewritten as
\begin{align}
    \lambda_A=\frac{|g_F|^4k_{Fp}^4}{6\pi^3v^2_{Fp}v_*}T^2\bigg[\frac{1}{\Delta_n}\arctan{\frac{\Delta_n}{|\xi_*|}}-\frac{|\xi_*|}{\xi_*^2+\Delta_n^2}\bigg].
\end{align}
In our low-proton-fraction $npe$ matter, there is still a significant gap between $2k_{Fp}$ and $k_{Fn}$, and $|\xi_*|\gg\Delta_n$ is thus satisfied. Therefore we have 
\begin{align}
    \frac{1}{2v_*\Delta_n}\arctan{\frac{\Delta_n}{|\xi_*|}}\simeq\frac{1}{2v_*|\xi_*|}-\frac{\Delta_n^2}{6v_*|\xi_*|^3}+\cdots,
\end{align}
\begin{align}
    \frac{|\xi_*|}{2v_*(|\xi_*|^2+\Delta_n^2)}\simeq\frac{1}{2v_*|\xi_*|}-\frac{\Delta_n^2}{2v_*|\xi_*|^3}+\cdots.
\end{align}
We can then obtain the asymptotic expression of $\lambda_A$ in our case:
\begin{align}
    \lambda_A\simeq \frac{|g_F|^4k_{Fp}^4}{9\pi^3v^2_{Fp}v_*}\frac{\Delta_n^2 T^2}{|\xi_*|^3}.
\end{align}
According to their asymptotic behavior, we can see that $\lambda_{\rm qp}$ decreases much more rapidly than $\lambda_{A}$ does when temperature becomes lower, which is consistent with the numerical results shown in Fig.~{\ref{fig:num-lambda-T}}. This indicates that the Andreev contribution may becomes more important than the quasiparticle one and even dominates in very low temperature, e.g. $T\sim 0.1\,{\rm MeV}$.

\section{Higher-order beta-tunneling currents other than Andreev reflection}

In the main text, we focused on the quasiparticle beta-conversion rate
\(\Gamma_{\rm qp}\) and the Andreev beta-conversion rate \(\Gamma_A\).
For completeness, we summarize here the remaining fourth-order tunneling
contributions, which are the beta-decay analogues of
\(I_1^{(3)}\) and \(I_2^{(3)}\) in the Rabi-coupled normal-superfluid
junction of Ref.~[16].

The fourth-order current contains three types of contributions:
\begin{equation}
  \Gamma^{(3)}
  =
  \Gamma_1^{(3)}
  +
  \Gamma_2^{(3)}
  +
  \Gamma_A .
\end{equation}
Here \(\Gamma_A\) is the Andreev beta-conversion rate discussed in the
main text, while \(\Gamma_1^{(3)}\) and \(\Gamma_2^{(3)}\) are the
counterparts of the higher-order non-Andreev tunneling currents
\(I_1^{(3)}\) and \(I_2^{(3)}\) of Ref.~[16].

The first contribution, \(\Gamma_1^{(3)}\), is the nonlinear
quasiparticle-tunneling correction.  In the present notation it is
written as
\begin{align}
  \Gamma_1^{(3)}
  =
  \frac{|g_F|^4}{16}
  \sum_{\bm{k}_p,\bm{k}_e}
  \sum_{\bm{k}'_p,\bm{k}'_e}
  \sum_{\sigma,\sigma_e,\sigma'_e}
  \delta_{\bm{k}_p+\bm{k}_e,\bm{k}'_p+\bm{k}'_e}
  \int
  \frac{d\omega_n}{2\pi}
  \frac{d\omega_p}{2\pi}
  \frac{d\omega_e}{2\pi}
  \frac{d\omega'_p}{2\pi}
  \frac{d\omega'_e}{2\pi}
  \nonumber\\
  \times
  \left[
    A_n(\bm{k}_p+\bm{k}_e,\omega_n)
  \right]^2
  A_p(\bm{k}_p,\omega_p)
  A_e(\bm{k}_e,\omega_e)
  A_p(\bm{k}'_p,\omega'_p)
  A_e(\bm{k}'_e,\omega'_e)
  \nonumber\\
  \times
  2\pi\delta(
    \omega_n-\omega_p-\omega_e+\delta\mu_\beta
  )
  \,
  2\pi\delta(
    \omega_n-\omega'_p-\omega'_e+\delta\mu_\beta
  )
  \nonumber\\
  \times
  \Bigl[
    f(\omega_n)
    \{1-f(\omega_p)\}
    \{1-f(\omega_e)\}
    \{1-f(\omega'_p)\}
    \{1-f(\omega'_e)\}
  \nonumber\\
  \hspace{2.0cm}
    -
    \{1-f(\omega_n)\}
    f(\omega_p)
    f(\omega_e)
    f(\omega'_p)
    f(\omega'_e)
  \Bigr] .
  \label{eq:Gamma_1_3_beta}
\end{align}
This term involves only the normal neutron spectral function
$A_n$. 
Therefore, in a neutron superfluid,
\(\Gamma_1^{(3)}\) is suppressed at low temperatures by the neutron
quasiparticle gap, in the same way as the leading quasiparticle rate
\(\Gamma_{\rm qp}\).

The second non-Andreev contribution, \(\Gamma_2^{(3)}\), contains the
anomalous neutron propagator but does not describe the conversion of an
incident \(pe\) particle channel into a \(pe\) hole channel.  It is the
analogue of \(I_2^{(3)}\) in Ref.~[16].  In the present
notation, we obtain
\begin{align}
  \Gamma_2^{(3)}
  =
 |g_F|^4
  \sum_{\bm{k}_p,\bm{k}_e}
  \sum_{\bm{k}'_p,\bm{k}'_e}
  \sum_{\sigma,\sigma_e,\sigma'_e}
  \delta_{\bm{k}_p+\bm{k}_e,\,-\bm{k}'_p-\bm{k}'_e}
  \int
  \frac{d\omega}{2\pi}
  \frac{d\omega_p}{2\pi}
  \frac{d\omega_e}{2\pi}
  \frac{d\omega'_p}{2\pi}
  \frac{d\omega'_e}{2\pi}
  \nonumber\\
  \times
  \left[
    {\rm Im}\,F_n(\bm{k}_p+\bm{k}_e,\omega)
  \right]^2
  A_p(\bm{k}_p,\omega_p)
  A_e(\bm{k}_e,\omega_e)
  A_p(-\bm{k}'_p,\omega'_p)
  A_e(-\bm{k}'_e,\omega'_e)
  \nonumber\\
  \times
  2\pi\delta(
    \omega-\omega_p-\omega_e+\delta\mu_\beta
  )
  \,
  2\pi\delta(
    -\omega-\omega'_p-\omega'_e+\delta\mu_\beta
  )
  \nonumber\\
  \times
  \Bigl[
    f(\omega)
    \{1-f(\omega_p)\}
    \{1-f(\omega_e)\}
    f(\omega'_p)
    f(\omega'_e)
  \nonumber\\
  \hspace{2.0cm}
    -
    \{1-f(\omega)\}
    f(\omega_p)
    f(\omega_e)
    \{1-f(\omega'_p)\}
    \{1-f(\omega'_e)\}
  \Bigr] .
  \label{eq:Gamma_2_3_beta}
\end{align}
The appearance of \([{\rm Im}\,F_n]^2\), rather than \(|F_n|^2\), is the
direct counterpart of the distinction between \(I_2^{(3)}\) and the
Andreev current in the Rabi-coupled problem.  In particular,
\(\Gamma_2^{(3)}\) is sensitive to the on-shell anomalous spectral weight
of the neutron superfluid.

At low temperatures, the nonlinear quasiparticle contribution
\(\Gamma_1^{(3)}\) is suppressed by the neutron quasiparticle gap because
it contains the normal neutron spectral function \(A_n\).  The terms
\(\Gamma_2^{(3)}\) and \(\Gamma_A\), on the other hand, involve the
anomalous propagator \(F_n\).  Among them, \(\Gamma_A\) is the genuine
Andreev beta-conversion contribution, since it describes the conversion
of a normal \(pe\) particle channel into a \(pe\) hole channel through
the neutron condensate.

\end{widetext}

\end{document}